\documentclass[twocolumn]{aastex631}
\usepackage{amsmath}
\usepackage{float}
\usepackage{xcolor}

\begin{document}

\title{ Mining the GALAH data I: Study of five Super lithium-rich metal-poor giants.}

\author[0000-0002-8519-7353]{Antony Susmitha}
\affiliation{Indian Institute of Astrophysics\\
Koramangala II block 560036 \\
Bengaluru, KA, India}

\author[0000-0001-9246-9743]{Anohita Mallick}
\affiliation{Indian Institute of Astrophysics\\
Koramangala II block 560036 \\
Bengaluru, KA, India}
\author[0000-0002-4282-605X]{Bacham E. Reddy}
\affiliation{Indian Institute of Astrophysics\\
Koramangala II block 560036 \\
Bengaluru, KA, India}

\begin{abstract}
The presence of a large amount of Li in giants is still a mystery. Most of the super Li-rich giants reported in recent studies are in the solar metallicity regime. Here, we study the five metal-poor super Li-rich giants (SLRs) from GALAH Data Release 3 with their [Fe/H] ranging from -1.35 to -2.38 with lithium abundance of A(Li) $\geq$ 3.4~dex.   The asteroseismic analysis reveals that none are on the red giant branch. The average period spacing ($\Delta P$ ) values indicate giants are in the core He-burning phase. All of them are low-mass giants (M $<$ 1.5M$_{\odot}$). The location in the  HR diagram suggests one of them is in the red clump phase, and interestingly, the other four are much brighter and coincide with the early AGB phase. The abundance analysis reveals that C, O, Na, Ba, and Eu are normal for giants of respective metallicities and evolutionary phases. Further, we didn't find any strong evidence for the presence of dust in the form of infrared excess or binarity from the available radial velocity data. We discussed a few scenarios for the existence of SLRs
 at higher luminosity, including past merger events. The findings will help to understand the production and evolution of Li among giants, in particular, during and the post-red clump phase.
\end{abstract}


\section{Introduction} \label{sec:intro}
Lithium is expected to deplete as stars evolve off the main sequence and climb up towards the tip of the red giant branch (RGB). The models predict Li abundance (A(Li)= $log N(Li)/log N(H) + 12$), post first dredge-up, not more than A(Li) = 1.5 to 1.8~dex depending on mass \citep{iben1967ApJ, gratton2000, lind2009}. A few among them have A(Li) exceeding the initial abundance of ISM, A(Li) = 3.2 dex \citep{knauth2003ApJ} with which stars have formed. The high A(Li) puzzle in red giants has been there for more than four decades since its discovery in 1982 \citep{luck1982, wallerstein1982}. 

In recent years, significant advances have been made in understanding the origin of high A(Li) in giants. This is mainly due to large data sets from spectroscopic surveys like Large Sky Area Multi-Object Fibre Spectroscopic Telescope (LAMOST) and Galactic Archaeology with HERMES (GALAH) and time-resolved photometry from space missions like Kepler. A large number of spectra helped increase the Li-rich sample by many folds, and the asteroseismic data from the Kepler mission helped resolve the stars' evolutionary phase in the Hertzsprung-Russel Diagram (HRD).  These studies suggest that all the Li-rich giants are in the He-core burning phase of the red clump (RC) region \citep{deepak2019MNRAS, singh2019ApJ}. The study by \citet{kumar2020} demonstrated that giants ascending the RGB only deplete Li and reach A(Li) as low as $-$0.9 dex towards the RGB tip. Multiple pieces of evidence were put forward that the high Li origin lies during the He-flash episode, which terminates the RGB phase of evolution. They argued that the He-flash is only the main stellar episode between the RGB phase's end and the red clump's beginning. Hence, the He-flash holds the key to the origin of high Li in RCs. In a novel study,  \citet{Singh2021ApJ} have shown that the A(Li) in RC giants and the gravity mode period spacing ($\Pi$p) are correlated. In that, the giants with high A(Li), relatively younger RCs,  have low $\Pi$p values compared to giants with normal A(Li), relatively older RCs. They concluded that A(Li) enrichment probably occurred within 2M years since the He-ignition began at the RGB tip. The study also demonstrated that Li-richness is a transient phenomenon. Further evidence was put forward by \citet{Mallick2023} in which they studied Li abundance in samples of Low ($\leq$2M$_{\odot}$) and high mass ($>$ 2M$_{\odot}$) giants. Interestingly, they found no Li-rich giants among high-mass stars, suggesting He-flash is the most likely cause for Li enrichment as high-mass giants are not expected to undergo He-flash, but low-mass giants do.  
It is reasonable to believe that Li enrichment occurred during the He-flash. However, it is not very clear how the Li produced in the interiors was brought to the surface, though a few mechanisms have been proposed \citep{cameron1971ApJ,fekel1993, sackmann1999, charbonnel2000, denissenkov2003apj, kumar2011ApJ,  lattanzio2015MNRAS}. Most of the Li-rich giants in the recent literature are in the metal-rich regime, i.e. [Fe/H] $\geq$ $-$1.0~dex.

Not many Li-rich giants were found among metal-poor giants. Only a handful of stars were detected in the metal-poor regime \citep{ruchti2011lirich, hainingli2018, casey2019} whose evolutionary phase was not explicitly constrained. The question is whether the mechanism that drives the Li enhancement in metal-rich giants is the same for the metal-poor giants.  Increasing the Li-rich metal-poor giants and resolving their evolutionary phase may provide further constraints on Li production, dredge-up process and its evolution. 

Here, we report results on five super Li-rich giants found while searching among a sample of metal-poor giants in the GALAH DR3 for which asteroseismic data is available.

\section{Sample selection}
\label{sec:sampleselection}
\begin{deluxetable*}{cccccccc}
   \caption{The basic data of the five SLR giants from GALAH DR3. \label{tab:radec}} 
    \tablehead{\colhead{Object name } & \colhead{RA} & \colhead{DEC} & \colhead{Vmag} & \colhead{A(Li)} & \colhead{log(L/L$_{\odot}$)} & \colhead{RV$_{GALAH}$} & \colhead{RV$_{GaiaDR3}$ } \\
      \colhead{}&\colhead{  (hh:mm:ss)} & \colhead{ (dd:mm:ss)} &\colhead{} & \colhead{(dex)}& \colhead{} & \colhead{(km/s)} & \colhead{(km/s)}
    }
 \startdata
 UCAC4 253-045343 & 10 25 28.84 &-39 26 01.09&12.6 &3.74 &3.01 & 171.3 $\pm$ 0.13 & 170.8 $\pm$ 0.24\\
UCAC4 099-098976 & 23 07 10.10 & -70 18 55.61&12.8 & 4.08& 1.70 &183.0 $\pm$ 0.31 & 182.3 $\pm$ 1.51 \\
 UCAC4 212-183136 & 20 29 16.13 & -47 41 51.51 &13.8 &4.23 & 2.71 & 5.9 $\pm$ 0.23 & 6.2 $\pm$ 1.53\\
UCAC4 308-077592 & 14 31 09.66 & -28 29 45.42 & 13.3 &4.80& 3.23 & -115.7 $\pm$ 0.13 & -115.0 $\pm$ 1.06\\
TYC 7262-250-1 & 13 09 32.26 & -37 09 17.79&11.5 & 4.54&3.00 & 30.7 $\pm$ 0.10 & 30.3 $\pm$ 0.34\\
 \enddata
\end{deluxetable*}
We utilised the recent data release (DR3) of the GALAH survey \citep{buder2021}  to study the lithium (Li) abundance in metal-poor giants.  The GALAH survey employs the High Efficiency and Resolution Multi-Element Spectrograph (HERMES, \citet{sheinis2015}) mounted on the 3.9 m Anglo Australian Telescope (AAT). HERMES provides high-resolution ( R $\sim$ 28000) optical spectra in four wavelength windows (4713–4903 \AA, 5648–5873 \AA, 6478–6737 \AA, and 7585–7887 \AA) covering the spectral features of up to 30 elements, including Li. We used specific selection criteria provided in the GALAH catalogue as bit-flags to ensure the quality of the data and accuracy in stellar parameter estimation. These criteria include {\tt \string flag{\_}sp = 0}, {\tt \string flag{\_}fe{\_h} = 0}, and {\tt \string flag{\_}Li{\_}fe = 0}. The bit-flag = '0' indicates that abundances are reliable and no problems were detected in determining the stellar parameters, iron and Li abundances. 
Further, since we are only focusing  on metal-poor giants, we applied the criteria {\tt \string logg $<$ 3.0} and {\tt \string [Fe/H] $\leq $ -1.0}, which resulted in a sample of 1038 metal-poor giant stars from GALAH DR3 survey.  Among the selected sample stars, we found five SLR giants with very high Li abundances, higher than the present interstellar medium value (A(Li) $\sim$ 3.2). 
We imposed no other selection criteria to restrict the samples, such as mass or Signal Noise Ratio. The spectra of five giants were downloaded from the GALAH survey site for analysis of the spectra and to search for additional peculiarities that may be present.  The details of the five giants are provided in  Table \ref{tab:radec}. Note \citet{martell2021} listed all five giants as Li-rich giants in their catalogue.

\begin{figure}[ht!]
    \centering
    \includegraphics[width=\columnwidth]{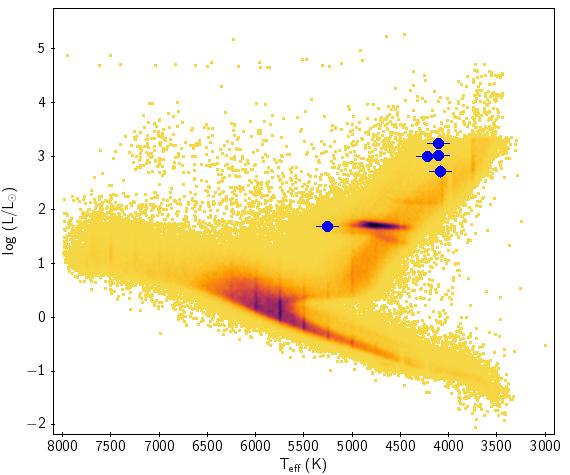}
    \caption{The HR diagram showing the five metal-poor SLRs along with stars in the background from the GALAH DR3 (yellow points). Note the well-defined RGB, luminosity bump and the red clump. One of the five SLRs is at the RC and the remaining four are at or close to the early AGB.}
    \label{fig:hrdiagram}
\end{figure}
\section{Stellar parameters and elemental abundances} \label{sec:stparam}
We found minor differences between the values by comparing the temperatures derived from photometric colours and those listed in the GALAH catalogue.  Since the Li abundance is derived from the resonant line at 6707 \AA\, which is very temperature sensitive, any variation in the temperature can lead to potential uncertainties in Li abundance values.  To address this issue and ensure the reliability of the values adopted from the catalogue data, we re-derived the stellar parameters and elemental abundances by analysing the spectra.
We used version 12 of the spectral synthesis code TURBOSPECTRUM developed by \citet{Plez2012ascl} to derive abundances and stellar
parameters. We used the stellar atmospheric models by \citet{meszaros2012aj} in which the ATLAS9 and MARCS codes were modified with an updated H$_{2}$O line list and with a wide range of C and $\alpha$-enhancements. We have used only carbon normal models as none of our stars exhibit enhancement in carbon. One-dimensional local thermodynamic equilibrium (LTE) is assumed for all the species. We have applied the non-LTE corrections wherever they are available in the literature (see  Table \ref{tab:abundstparam}). \\
\\
We adopted the solar abundances from \citet{asplund2009araa}.
The line lists for atomic lines were assembled from the Vienna
Atomic Line Database (VALD) \citep{kupka1999aas}. Hyper-fine structure has been
accounted for transitions of elements Li, Ba, and Eu. For the case of molecular line lists, we used  C2 data from the Kurucz database (Kurucz 2007, 2009, 2013), and  CN data from \citet{plez-cohen2005aa}.
\subsection{Stellar parameters} \label{subsec:parameters}
\begin{deluxetable*}{cccccccccccc}
 \caption{Atmospheric parameters of the five SLR giants taken from the GALAH catalogue and derived in this study. Also, given is the $T_{\rm eff}$ derived from the photometry and listed in the Gaia. The uncertainties associated with the estimation of stellar parameters in this study are $\pm$150 K for Teff, $\pm$0.25 dex for log $g$, and $\pm$0.15 km/s for $\xi_{t}$. \label{tab:stparam} } 
 \tablehead{ 
 \colhead{} &\multicolumn{4}{c}{GALAH DR3} & \colhead{Teff} & \colhead{Teff} & \multicolumn{5}{c}{$This~ study$} \\
 \cline{2-5} \cline{8-12}
 \colhead{Object name} & \colhead{ Teff} & \colhead{log $g$} & \colhead{[Fe/H]} & \colhead{$\xi_{t}$} & \colhead{(V-K)} & \colhead{ $Gaia$} &\colhead{Teff}& \colhead{log $g$}& \colhead{[Fe/H]} & \colhead{$\xi_{t}$}& \colhead{$vsini$} \\
 \cline{11-12}
 \colhead{} & \colhead{ (K) } & \colhead{(dex)} & \colhead{(dex)} & \colhead{(km/s)} & \colhead{(K)} & \colhead{ (K)} &\colhead{(K)}& \colhead{(dex)}& \colhead{(dex)} & \multicolumn{2}{c}{(km/s)}
   }
   \startdata
   UCAC4 253-045343 & 4308$\pm$90 & 1.06$\pm$0.31& -1.26$\pm$0.07&1.6 &4025 & 4327 & 4100&  1.00&  -1.43$\pm$0.14&  1.4& 6.0 \\
    UCAC4 099-098976 &  5263$\pm$101& 2.49$\pm$0.22& -1.46$\pm$0.12& 1.1 &5205 & 5375 & 5260&  2.49&  -1.46$\pm$0.13&  2.1& 19.1\\
  UCAC4 212-183136  &4505 $\pm153$& 1.49$\pm$0.36& -1.40$\pm$0.13& 1.4 &4091 & 4318 &4080&  0.10& -2.38$\pm$0.15& 2.7& 13.8 \\
 UCAC4 308-077592 & 4210 $\pm$89& 0.60$\pm$0.45& -1.67$\pm$0.07& 1.6&3923 & 4206 & 4100&  0.60& -1.86$\pm$0.13& 1.7& 6.2\\
 TYC 7262-250-1 &4218$\pm$73& 0.82$\pm$0.34& -1.31$\pm$0.05&1.6 &4193 & 4121 & 4217& 0.82& -1.36$\pm$0.12& 1.7& 4.3\\
   \enddata
\end{deluxetable*}
 We have measured the equivalent width (EW) of unblended neutral and singly ionised Fe lines to derive the stellar parameters.   EWs were measured by fitting Gaussian profiles to the features using {\it splot} task in PyRAF \footnote{PYRAF is a product of the Space Telescope Science Institute, 
operated by AURA for NASA.}. We have considered only those lines  whose EW is less than
100 m\AA\, since they are on the linear part of the curve-of-growth and are relatively insensitive to the choice of microturbulence \citep{mucciarelli2011aa}.
The uncertainties in the
measurements were determined using the revised Cayrel formula
\citep{cayrel1988iaus, battaglia2008mnras}.  
We have interpolated the
models from the grid of atmospheric models provided by \citet{meszaros2012aj} to obtain models of  specific stellar parameters.
The $T_{eff}$ was derived  by forcing the abundances of Fe~I
lines giving the same abundance irrespective of the line's lower excitation potential (LEP). We estimated an error of 150 K on the spectroscopic temperature by ensuring that any alteration in the slope of iron abundances derived from neutral lines with excitation potential remained within one standard deviation of the Fe~I line abundances. The usual technique to derive the surface gravity (log {\it g}) assumes ionisation equilibrium between Fe~I and Fe~II lines. Since we have only one clean  Fe~II  line detected in the GALAH spectra, we adopted the log {\it g} value when the Fe~II abundance is within the 1$\sigma$ of the average Fe~I abundances. We assumed an error of 0.25 dex for the estimation of log {\it g}.
Microturbulent velocity ($\xi_{t}$) is calculated by establishing the abundance of Fe~I to be independent of reduced equivalent width (log(EW/$\lambda$)). The error in the estimation of $\xi_{t}$ is obtained when the Fe~I abundance did not change by more than 1$\sigma$  and the error is 0.15 km $^{-1}$. The method was iterated until we found a set of $T_{\rm eff}$, log $g$, and $\xi_{t}$ values for which we found no significant slope between abundances of Fe~I lines and their LEP values, measured EWs, and the abundances of neutral and ionised lines are equal.  The final value of Fe abundance obtained for the converged atmospheric parameters is taken as the metallicity of a star. The derived stellar parameters, the estimated errors in this study, and the values given in the GALAH catalogue are listed in Table \ref{tab:stparam} along with values of $T_{\rm eff}$ derived from the colour (V-K) and values given in the Gaia catalogue. Values of $T_{\rm eff}$ derived here agree very well with the values from Gaia and (V-K) but one could notice a significant difference of about 500K  with GALAH value for a giant UCAC4 212-183136. For the same star, we also found a large difference in $log g$. However, the $log g$ derived from the Gaia parallax agrees well with our value (see Table \ref{tab:luminMv}).  The differences between the values derived here and the values given in the GALAH DR3 catalogue may be attributed to the choice of model atmospheres and, to some extent method of analysis; pipeline versus manual analysis of star by star. Since the $T_{\rm eff}$ and $log g$ derived here are in good agreement with the values derived from Gaia parallax we adopt stellar parameters obtained in this study for abundance analysis. 
\subsection{Abundances}
\begin{figure*}[!ht]
    \centering
    \begin{tabular}{ccc}
         \includegraphics[width = 0.32\textwidth]{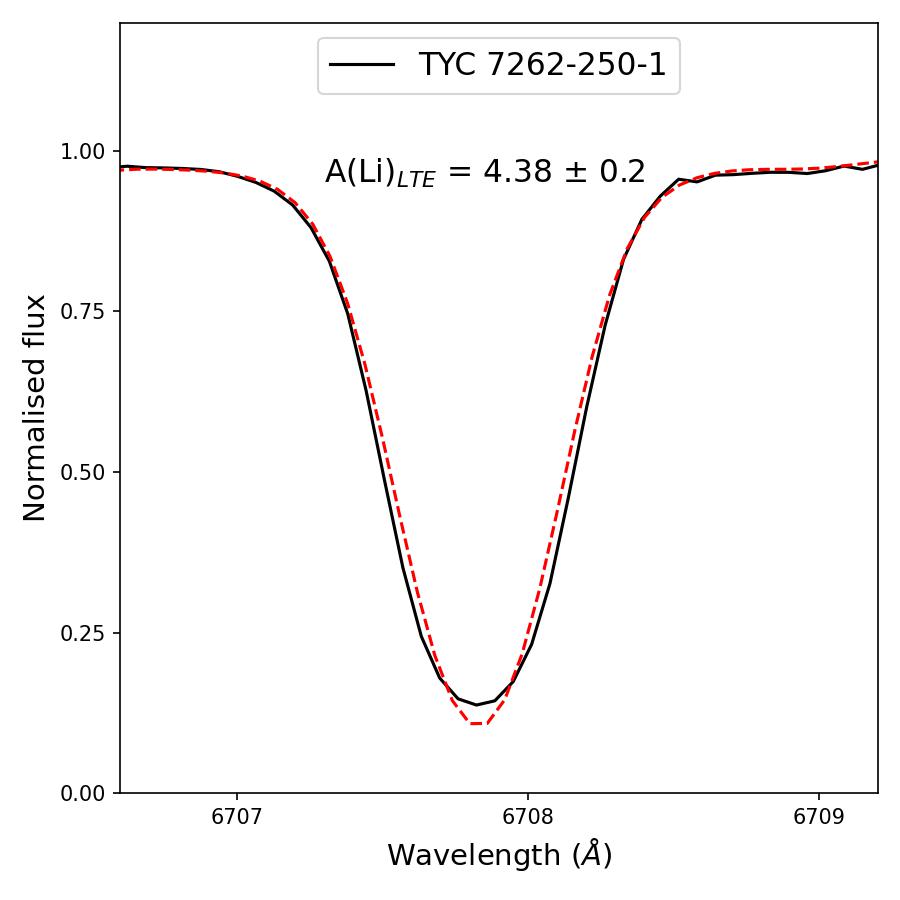} & \includegraphics[width=0.32\textwidth]{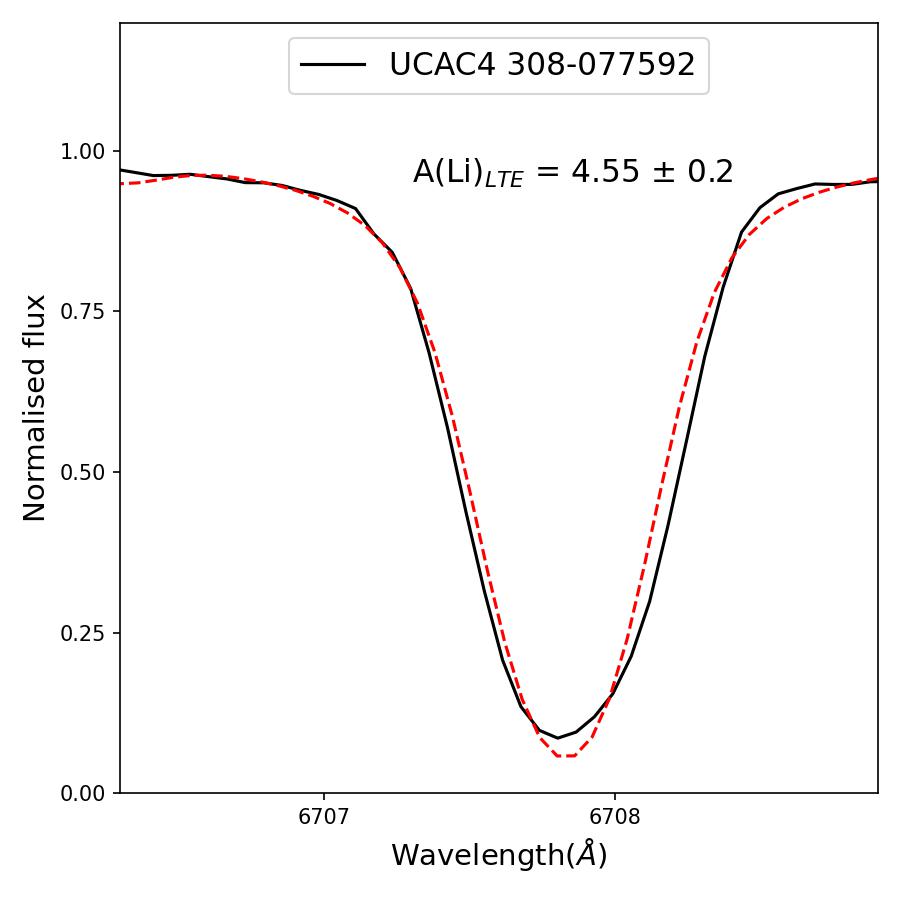} &
       \includegraphics[width=0.32\textwidth]{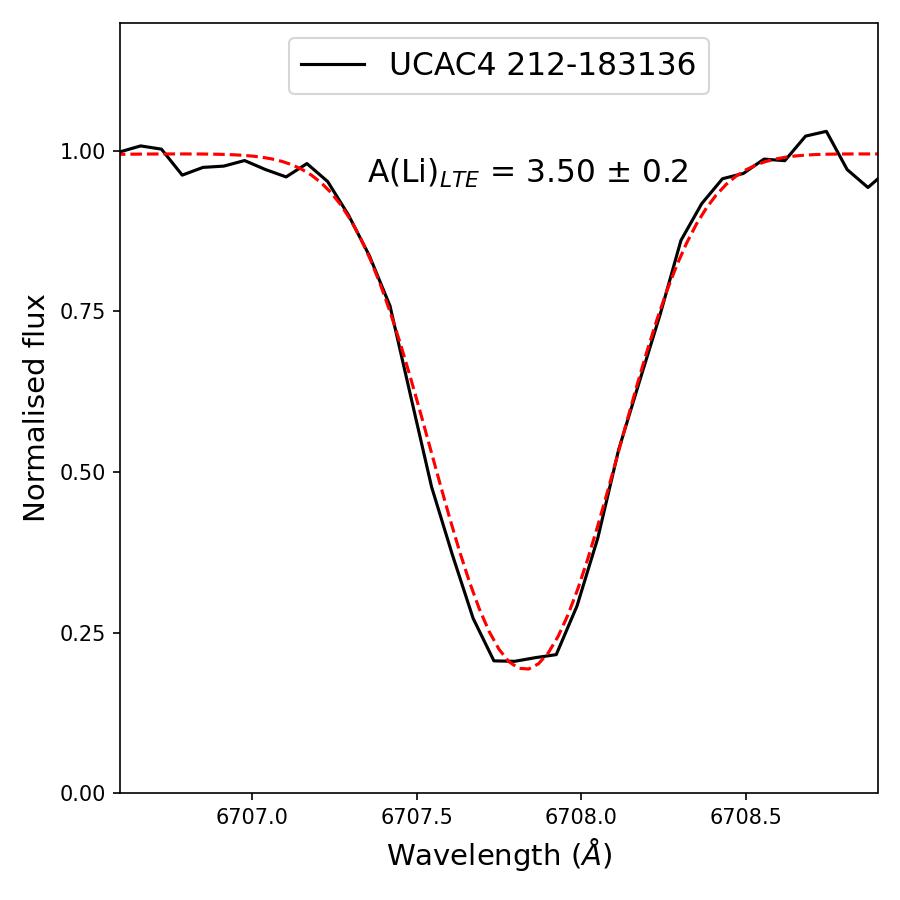} \\ \includegraphics[width=0.32\textwidth]{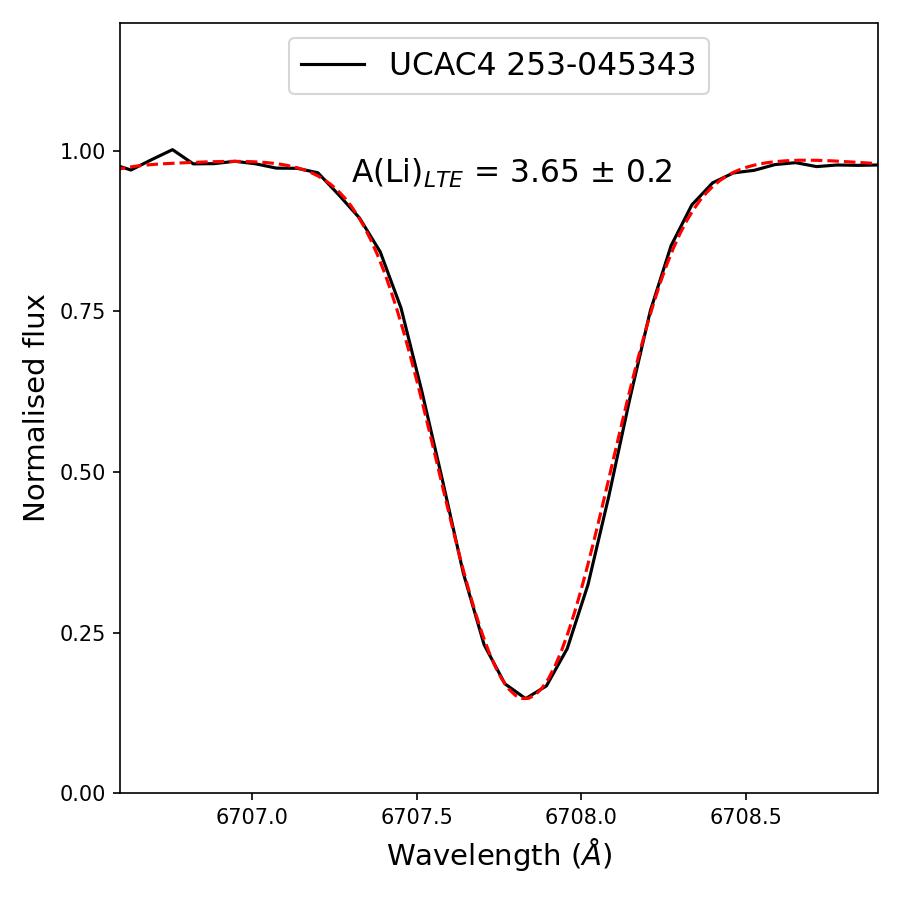}  &  \includegraphics[width = 0.32\textwidth]{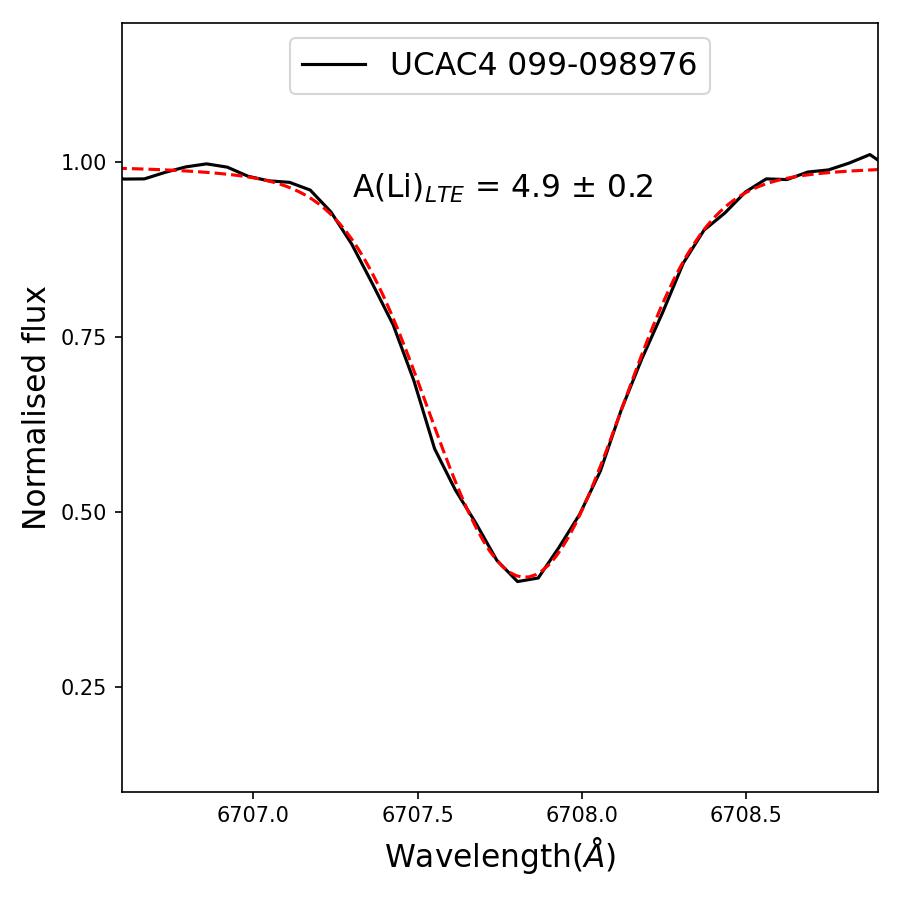} & \\
    \end{tabular}
    \caption{The Li line profiles at 6707 \AA\ of the five SLR stars compared with the best fit synthetic spectra (dashed line).
    }
    \label{fig:li6707}
\end{figure*}

\begin{deluxetable*}{cccccccc}
    \caption{The elemental abundances of the SLR stars in this study.\label{tab:abundstparam}}
    \tablehead{
\colhead{Object name} &\colhead{ A(Li)$_{LTE}$} & \colhead{A(Li)$_{NLTE}$} & \colhead{[C/Fe]} & \colhead{[O/Fe]} &\colhead{ [Na/Fe]} & \colhead{ [Ba/Fe]} &\colhead{ [Eu/Fe] } \\
\colhead{} &\colhead{$\pm$0.20}& \colhead{$\pm$0.20} &  \colhead{$\pm$0.20} & \colhead{$\pm$0.20} &\colhead{$\pm$0.20} & \colhead{$\pm$0.20} &\colhead{$\pm$0.30}
}
    \startdata
UCAC4 253-045343 & 3.65 & 3.60 & -0.14 & 1.03 & -0.27 & 0.56 & 0.75 \\ 
UCAC4 099-098976 & 4.90 & 4.21 & $<$ -0.15 & 0.96 & -0.24 & -0.02 & $<$ 0.20 \\
UCAC4 212-183136 & 3.50 & 3.45 & 0.08 & $-$ & 0.58 & -0.38 & 0.86 \\
UCAC4 308-077592 & 4.45 & 4.42 & -0.03 & 1.16 & -0.24 & -0.13 & 0.45 \\
TYC 7262-250-1 & 4.38 & 4.31 & -0.19 & 0.55 & -0.05 & -0.09 & 0.59 \\
\enddata
\end{deluxetable*}
Due to discrepancies between the stellar parameters from the GALAH data and our values, we reevaluated the abundance of lithium and the elements of interest for this study, which include C, O, Na, Ba, and Eu. We used the resonance line at 6707.8 \AA\ to derive the Li abundance and assumed that all the Li present was from the $^{7}$Li isotope. The other Li I line, 6103\AA,  is not in the spectral range covered by the GALAH survey. We have obtained the NLTE corrections for the derived Li abundances from \citet{lind2009LiNLTE} , and the corrected values are given in Table \ref{tab:abundstparam}. We also used the 3D NLTE correction code {\tt \string BREIDABLIK} \citep{wang2021MNRAS} to obtain the 3D NLTE Li abundance. However, the code issued a warning saying the stellar parameters and LTE Li abundances of all stars in this study are  outside the grid utilized in {\tt \string BREIDABLIK} and the corrections may not be reliable.  Nonetheless, according to the grid from \citet{lind2009LiNLTE}, only the Li abundance in UCAC4 099-098976 falls outside the grid where we performed a linear extrapolation of the grid to obtain the NLTE correction for the star. One could see that the NLTE corrections from \citet{lind2009LiNLTE} are very small ($\sim$ -0.05) for the cool stars in our sample whereas, for UCAC4 099-098976, which is relatively hotter, the NLTE correction is $\sim$ -0.7 (1D NLTE(A(Li) - 1D LTE A(Li))). Such larger correction were also reported previously  by \citet{hainingli2018} and  \citet{sanna2020aa}.  May be the higher correction is  needed as the star is  relatively hotter and metal poor in which large over ionisation takes place. 
The plots for the best fit with their final abundance values are shown in Fig. {\ref{fig:li6707}}. \\

The GALAH spectra also cover the spectral features O~I triplet near 7774 Å, Ba~II at 5853.7~\AA\ and 6496.9 \AA\, and Eu~II at 6645.1~\AA. Using the spectral synthesis method, we derived the abundances of these elements, and the values are given in Table \ref{tab:abundstparam}. To determine the abundance of C, we examined the C$_{2}$ molecular band-head feature at 4737 \AA\, which is quite weak or absent. Constraining the $^{12}$C/$^{13}$C isotopic ratio was particularly challenging, as all our stars are deficient in carbon. Also, the $^{13}$C$_{2}$ feature at 4744 \AA\ is too faint to derive the abundance. Additionally, the usual $^{13}$CN features (7990 - 8040 \AA) used for deriving the $^{13}$C abundance are not covered by the GALAH spectrograph.
\\
\\
All the stars in our sample display depleted C abundance and elevated O  abundance. The O abundance was estimated using triplets, but it should be noted that these values have contributions from NLTE effects, and corrections for such effects are not available for cool stars similar to those studied here. Given this limitation, we can only confidently state that oxygen is enhanced. However, providing a quantitative measure of the enhancement after applying the NLTE correction is currently beyond the scope of this study.  
\subsubsection{Abundance errors}
The abundances of all the elements except Fe are derived using the spectral synthesis method. So, the uncertainties in the abundance values were estimated through the goodness of the least squares fit,
and the values are given in Table \ref{tab:abundstparam}. For deriving the Fe abundance, we used the method described in the section \ref{subsec:parameters} and the error due to contributions from
uncertainties in EWs and atomic parameters are listed as the 1$\sigma$ line-to-line scatter, $\sigma_{log \epsilon}$ in Table \ref{tab:stparam}. 
\section{Other observed properties}
\subsection{ The H\texorpdfstring{$_{\alpha}$}{alpha} profile and mass loss}
\begin{figure}[!ht]
    \centering
    \includegraphics[width=\columnwidth]{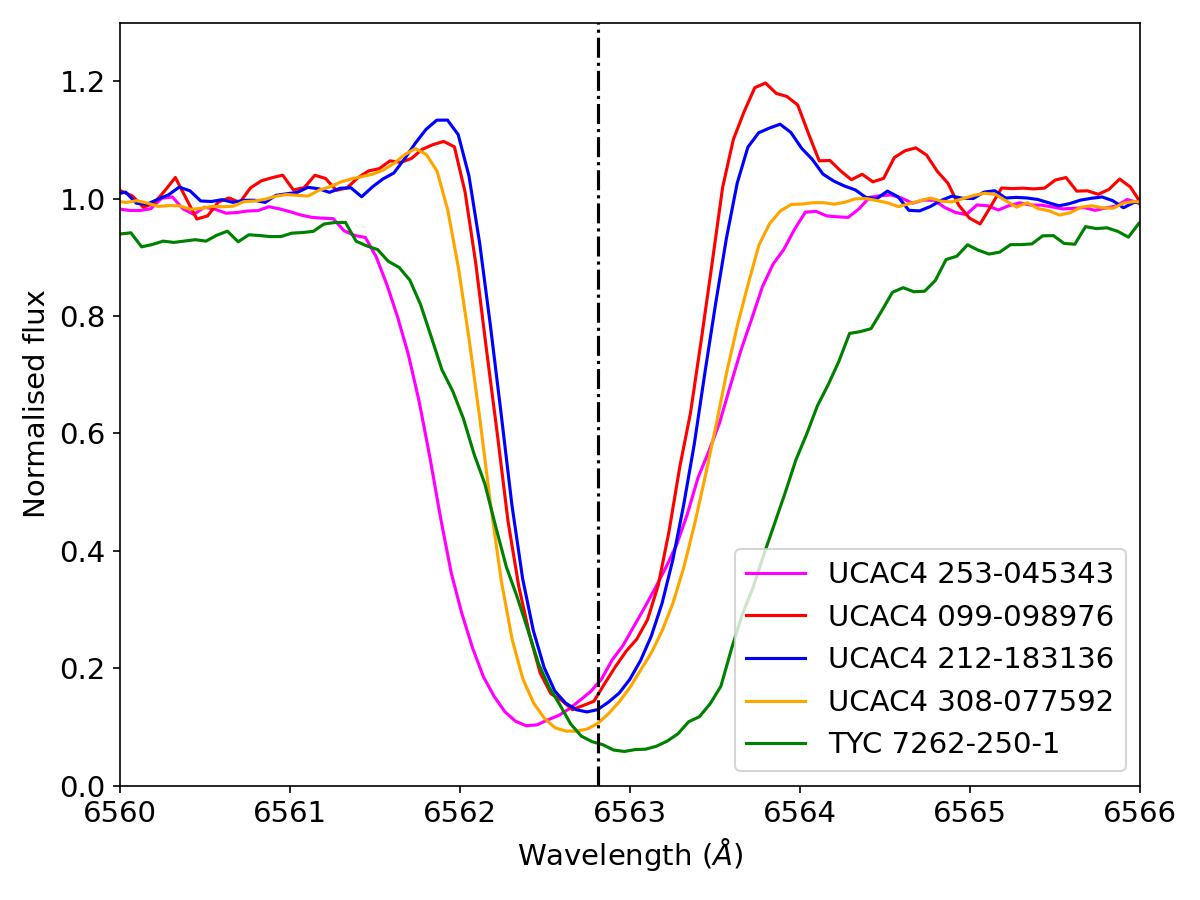}
    \caption{The H$_{\alpha}$ profile of five SLRs. Note, that all of them show asymmetry in the profiles. }
    \label{fig:halpha}
\end{figure}

The profile of H$_{\alpha}$ spectral line is an indicator of possible mass-loss/stellar activity present in the star and can be detected through asymmetries in the H$_{\alpha}$ line profile \citep{meszaros2009AJ}.  
 All five giants show distorted H$_{\alpha}$ profile, and three of them also show emission in the H$_{\alpha}$ wings as shown in Fig.  \ref{fig:halpha}. The emissions are also asymmetric. We also examined Ca-triplet features not covered in the GALAH spectra, using the {\it Gaia} DR3 data \citep{GaiaCollaboration2023aa}, and the feature seems normal. We looked for possible mass loss using $2MASS$ and WISE all-sky survey of infrared photometry. 

The W4 band magnitudes in the WISE catalogue have quality flags other than 'A'. Two of the samples have 'U' as the quality flag, which is expected for such fainter sources as the W4 band is proved to have very low sensitivity for reliable detection of fainter sources. So we did not calculate the W1-W4 value for those stars, but for the other three stars (UCAC4 253-045343, UCAC4 308-077592, TYC 7262-250-1), we have obtained the values for $W1-W4$ as  0.49, 0.64 and 0.25, respectively. The values are very low, suggesting no active mass loss. None of the giants shows any excess in the infrared colours (J, H, K and W1, W2, and W3 or W4), suggesting, at least, no hot dust component. We don't have far-IR colours for the sample to check if the stars have cold dust in case the mass loss occurred some time ago. 

As per the asymmetric H$_{\alpha}$ profiles, it is shown that the emission wings of the H$_{\alpha}$ line found in metal-poor stars can arise naturally from an extended, static chromosphere\citep{dupree1984ApJ, dupree1986araa}, and may not be solely the signature of mass-loss. So, the line asymmetries exhibited by these stars can be attributed to some disturbances in the stellar atmosphere, which would result in local mass flows rather than a steady-state mass loss.

\subsection{Rotational velocity}
\citet{charbonnel2010} suggests that the excess Li in some low-mass giants may be due to rotation-induced extra mixing. We derived the giants' projected rotational velocity to see if a correlation exists between Li abundance and the projected rotation velocity, $vsini$. The GALAH DR3 provided the overall line broadening parameter in the form of $vbroad$, which includes $vsini$, and thermal broadening macro-turbulence ($vmacro$). We disentangled the $vbroad$ into $vsini$ and $vmacro$ by adopting the relations between $vmacro$ and
$T_{\rm eff}$ for different luminosity classes as per the recipe given in \citet{hekker2007}. We have adopted the method described in \citet{ruchti2011lirich} to distinguish the luminosity class of each star. Since all the stars except UCAC4 099-098976  lie close to the RGB tip, we classified them as luminosity class II. For UCAC4 099-098976, we used the designation luminosity class III as its location in the HR diagram is closer to the RGB bump. The derived $vsini$ is given in the Table \ref{tab:abundstparam}. 
According to \citet{carney2008rotationVMacro}, the luminous giants with M$_{v}$ $<$-1.5 exhibit net rotation, and the projected rotation velocity can be larger than 3 km s$^{-1}$, a typical value for the upper RGB giants.

All the giants in this study show larger $vsini$ values than expected, particularly the two giants, UCAC4 099-098976, and UCAC4 212-183136. 
\citet{fekel1993} argued that during the convective mixing,  redistribution of angular momentum from the dredged-up material could induce increased rotation in stars, creating a dynamo resulting in chromospheric activity. Also, it could be the case where giants show higher rotation due to a sudden decrease in the size, i.e., radius due to He-flash and the subsequent transition to the red clump phase.
\subsection{Radial velocity}
The sample stars' multi-epoch radial velocity values are important to understanding any radial velocity variation observed. This can be used to distinguish whether any of our giants are a member of a binary system. We examined the radial velocity data from GALAH DR3 and Gaia DR3 \citep{katz2023aa} (see Table \ref{tab:radec}). The radial velocity from Gaia DR3 is obtained by median combining the individual epoch (transit) radial velocities. The individual epoch radial velocities are not available from Gaia DR3.  The RV data from the two catalogues match very well, within 1 km s$^{-1}$, though they are in two different epochs, indicating no radial velocity variations. 
We have also looked at the re-normalised unit weight error.
(RUWE) of the samples in the Gaia database. RUWE corresponds to the reduced chi-squared of the best-fitting 5-parameter single-body astrometric solution. This error coefficient can be used to identify possible non-single stars whose RUWE $>$ 1.4 \citep{lindegren2018, lindegren2021}. All the five giants have the RUWE $<$ 1.4, indicating a lack of binarity. However, for binary systems whose orbital period is larger than about  $\sim$ 1000 days, RUWE loses most of its efficiency in detecting binary systems, and RUWE parameter may not provide correct information about their binarity \citep{jorissen2019MmSAI}. Hence, even though the RUWE for the samples is $<$ 1.4, the radial velocity variation from a long-period binary companion cannot be ruled out.
The kinematic (U, V, W) data provided in the GALAH DR3 catalogue suggest that the four giants, including the RC star, belong to the Galactic halo component, whereas TYC 7262-250-1 belongs to the Galactic thick disk.

\begin{deluxetable}{ccccc}
\caption{The U,V,W space velocities of the SLR stars are given here. The last column corresponds to the total space velocity.\label{tab:spacevel} }
\tablehead{ 
    \colhead{Object name}& \colhead{ U }& \colhead{ V} & \colhead{W} & \colhead{Vtotal} \\
    \colhead{}& \colhead{ (km/s) }& \colhead{ (km/s)} & \colhead{(km/s)} & \colhead{(km/s)}}
    \startdata
UCAC4 253-045343  &-129.2 & -173.5 & 54.2 &  239.8 \\
UCAC4 099-098976 & -186.4& -328.6 & -158.1 & 429.6 \\
UCAC4 212-183136  & -46.8  & -259.0 &-24.4 & 282.0 \\
UCAC4 308-077592  & -74.0& -69.7 &-189.2 & 230.4 \\
TYC 7262-250-1 & -45.1& -14.2 &104.0 & 115.7  \\
\enddata
\end{deluxetable}
\section{The evolutionary phase } 
 
It is important to determine the evolutionary phase of Li-rich giants to constrain the origin of Li enhancement.  The evolutionary phase could be determined either using the star's location in the HR diagram combined with the evolutionary tracks or by using asteroseismic data. The latter turned out to be more reliable and is considered the gold standard for separating  RC giants of He-core burning from those of RGB giants with He-inert core. 
\label{sec:evolutionaryphase}
Recent studies concluded that super Li-rich giants are most likely to be RC stars rather than RGB stars \citep{deepak2019MNRAS, singh2019ApJ, kumar2020}. 
\subsection{Evolutionary tracks} \label{subsec:evolutionary phase}

GALAH DR3 provides a value-added catalogue (VAC) with luminosities calculated using the Bayesian Stellar Parameter Estimation code (BSTEP) from \citet{sharma2018}. BSTEP provides a Bayesian estimate of intrinsic stellar parameters from observed parameters by using the stellar isochrones. Although the luminosities of these stars are readily available from the VAC, due to the change in effective temperature from this study and from the GALAH DR3 catalogue (see Table \ref{tab:stparam}),  we have calculated the luminosity and log$g$ of these stars using parallaxes obtained from the Gaia DR3 and V band magnitude using the relation
\begin{equation}
    -2.5 log\frac{L}{L_{\odot}} = M_{V} + BC_{V}(T_{eff}) - Mbol_{\odot}
\end{equation}
and 
\begin{equation}
    log{\it g} = logg_{\odot}+log\frac{M}{M_{\odot}}+4log\frac{T_{eff}}{T_{eff_{\odot}}} - log\frac{L}{L_{\odot}}
\end{equation}
where the absolute magnitude in V band is calculated by 
\begin{equation}
    M_{V} = V + 5 - 5 log_{10} r - A_{V} 
\end{equation}
 and the temperature-dependent bolometric correction is calculated by 
 \begin{equation}
     BC_{V}(T_{eff}) = a+blogT_{eff}+c(logT_{eff})^2+d(logT_{eff})^3...
 \end{equation}
 Here,  {\it a, b, c, d } are the polynomial coefficients of the model function. Values of these coefficients are taken from \citet{torres2010aj}. 
The values of Mbol$_{\odot}$ = 4.74, log{\it g}$_{\odot}$ = 4.44 and T$_{eff_{\odot}}$ = 5772 K are adapted for the Sun. {\it r} is the distance to the star (in parsec), and A$_{V}$ is theinterstellar extinction in the V band. We have obtained the $A_{V}$ values from \citet{schlafly2011apj}.  We have used the  $T_{eff}$ derived in this study as the input $T_{eff}$ in the above equation, whereas the mass of the star is obtained from the asteroseismic analysis (refer section \ref{subsec:asteroseismicanalysis}.)
The final values are given in the Table \ref{tab:luminMv}. Within the estimated uncertainties, the log $g$ values derived from Gaia parallaxes agree well with those from GALAH DR3, except for the giant UCAC4 212-183136. For this star, the $log g$ from GALAH DR3 is more than 1~dex compared to the value derived in this study from the spectra and the Gaia parallax (see section \ref{subsec:parameters}).

\begin{deluxetable}{ccccc}
    \caption{The luminosity and log$g$ derived using Gaia parallaxes and V band magnitudes (see section \ref{subsec:evolutionary phase})of the SLR stars in this study. \label{tab:luminMv}}
    \tablehead{
        \colhead{ Object name} & \colhead{Av} & \colhead{$\pi$} & \colhead{$\log\left(\frac{L}{L_{\odot}}\right)$} & \colhead{log{\it g}} \\
        \colhead{} & \colhead{(mag)} & \colhead{(mas)} & \colhead{} & \colhead{(dex)}
        }
        \startdata
         UCAC4 253-045343  &   0.28 &  0.13 &     3.17$\pm$0.16 & 0.75$\pm$0.25 \\
         UCAC4 099-098976  &   0.08  & 0.37 &  1.73$\pm$0.12 & 2.56$\pm$0.16 \\
         UCAC4 212-183136  &  0.09 &   0.05  & 3.50$\pm$0.22 & 0.10$\pm$0.24 \\
         UCAC4 308-077592  &  0.24 &  0.05  & 3.66$\pm$0.18 & 0.33$\pm$0.21 \\
         TYC 7262-250-1 &  0.16  &  0.18  & 3.18$\pm$0.20 & 0.57$\pm$0.23  \\
    \enddata
\end{deluxetable}
The luminosity values derived here agree well with the values given in GALAH DR3 VAC within the estimated uncertainties except for one star, UCAC4 099-098976.  
The HR diagram of the data set from GALAH DR3 (See Fig.   \ref{fig:hrdiagram}) shows a well-defined main sequence, sub-giant and red giant branch. Also, one could notice two groups of stars on a red giant branch; one is identified as the luminosity bump and the other as a red clump or horizontal branch. Of the five giants in our sample, one overlaps with the red clump region, and the four seem to be much brighter and closer to the early AGB. 
\subsection{Asteroseismic analysis} \label{subsec:asteroseismicanalysis}
Asteroseismology has emerged as a valuable tool in distinguishing core helium-burning stars from hydrogen-shell-burning RGBs. The turbulent outer layers of RGBs display stochastically excited oscillations, which could be detected from high-precision time series data of long duration, typically facilitated by space missions such as Kepler \citep{Borucki2010}, CoROT \citep{Michel2008} and TESS \citep{Ricker2015}. Two distinct types of oscillations are seen in stars: pressure ($p$-mode) or acoustic mode, typically observed in the outer envelope of the star, and the gravity ($g$-mode) mode, predominantly found in the stellar core. 
Using the analysis of the oscillations associated with the $p$- and $g$-modes
in red giants, one could derive two characteristic parameters to distinguish giants with He-inert core, RGB giants, from those of He-core burning red clump giants. The two parameters are large frequency separation ($\Delta \nu$ of $P$-modes and the average period spacing ($\Delta$P) of mixed modes arising from coupling between the interior $g$-mode and the envelope $p$-mode oscillations. \\ 

For our analysis, we made use of data from NASA's Transiting Exoplanet Survey Satellite (TESS) from the Mikulski Archive for Space Telescopes (MAST) archive (\url{https://mast.stsci.edu/portal/Mashup/Clients/Mast/Portal.html}). We systematically removed points from the light curves with a quality flag greater than 0 or NaN flux values to ensure good data quality. Although this resulted in removing some data points, which can impact seismic analysis — the number of such excluded points remained minimal compared to the overall data size. To rectify these sudden jumps in measured flux, we applied correction using the TESS Asteroseismic Science Operations Center (TASOC) pipeline \citep{Handberg2021}. This pipeline employed a piece-wise cubic Hermite polynomial to interpolate the gaps, ensuring a more continuous and consistent light curve. Finally, all corrected light curves for a single object were stitched together using the lightkurve package \citep{2018Lightkurve}. The resultant light curves were converted to periodograms for seismic analysis using the Lomb-Scargle periodogram technique. The power density spectrum (PSD) is subjected to background removal using a log-median filter. The PSD is divided by the estimated background noise to create a "Signal-to-noise spectrum". \par

\begin{figure*}[!ht]
    \centering
    \begin{tabular}{ccc}
         \includegraphics[width=8cm]{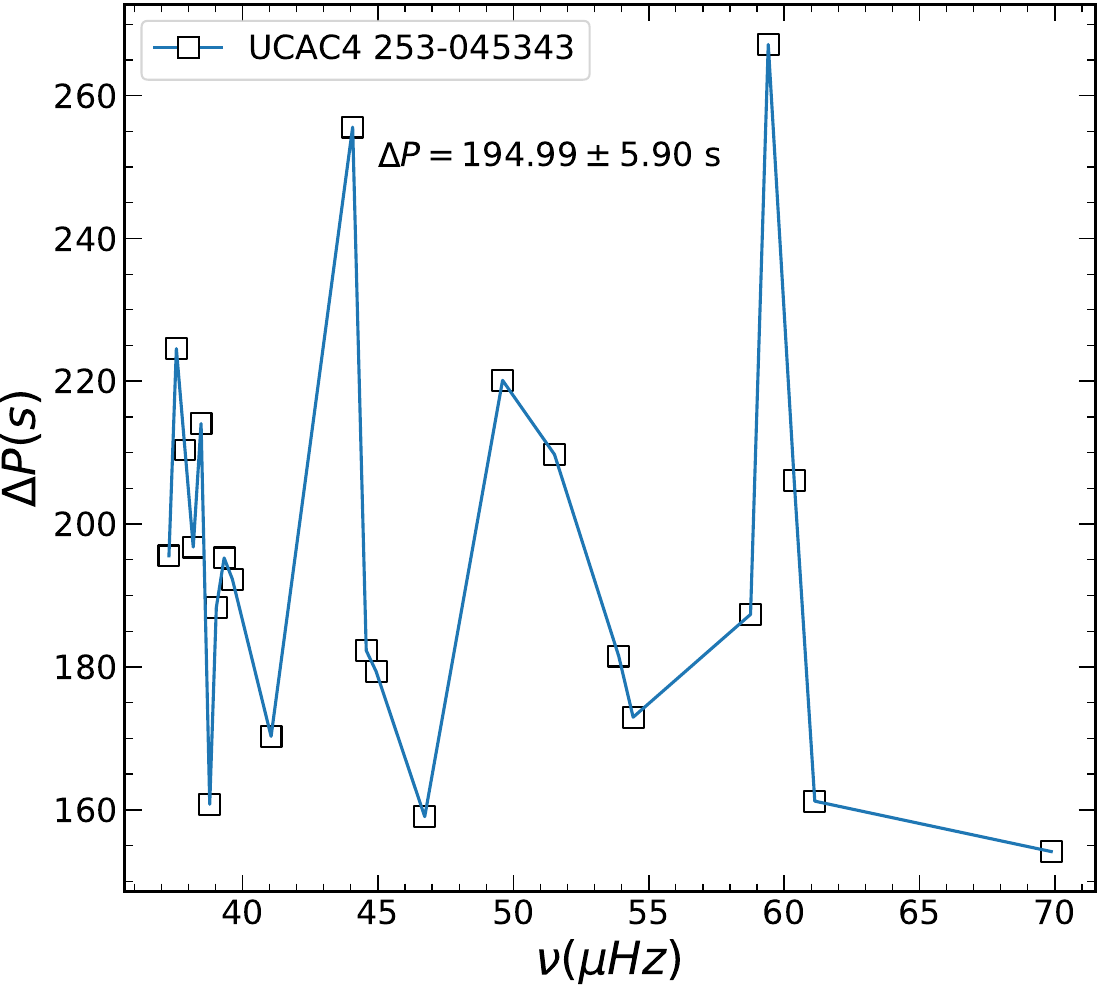} &  \includegraphics[width=8cm]{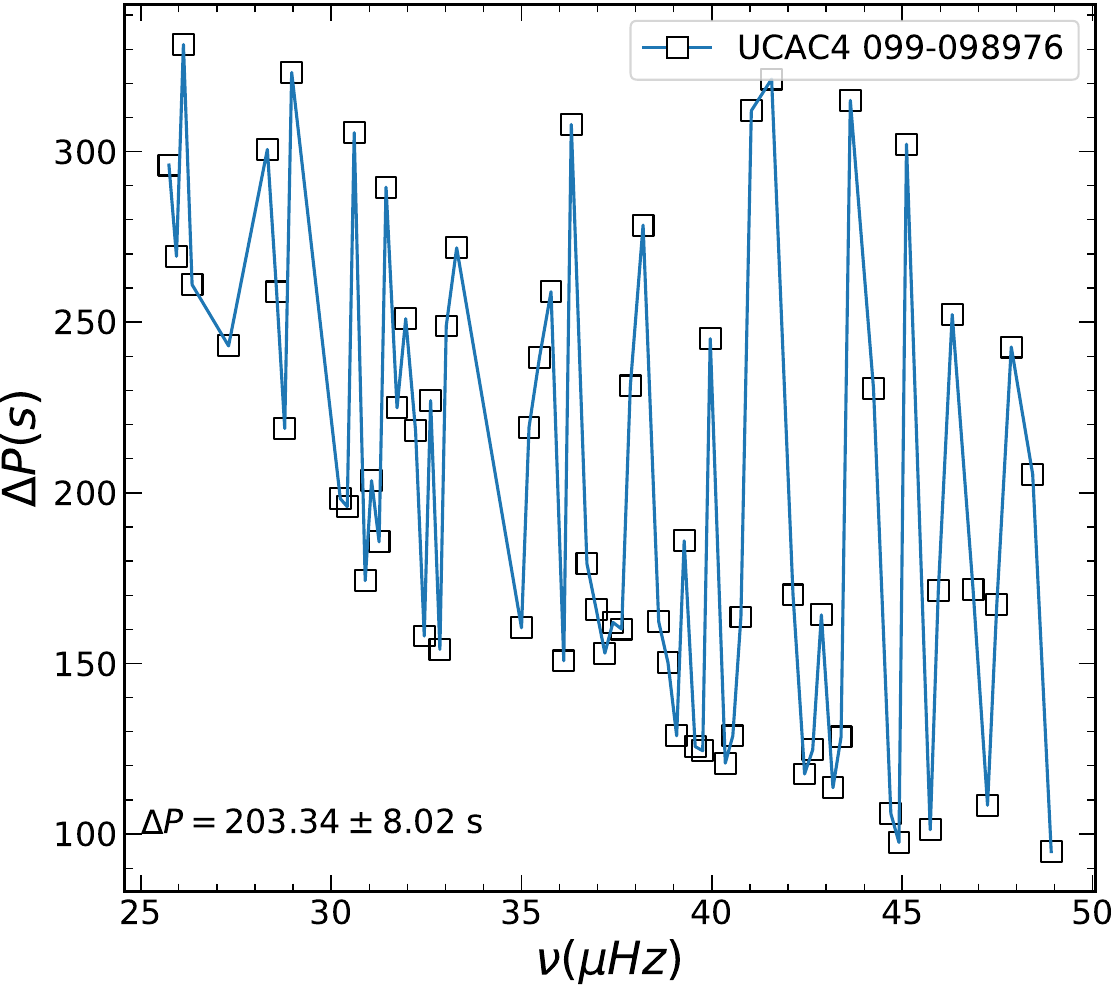}\\ 
         \includegraphics[width=8cm]{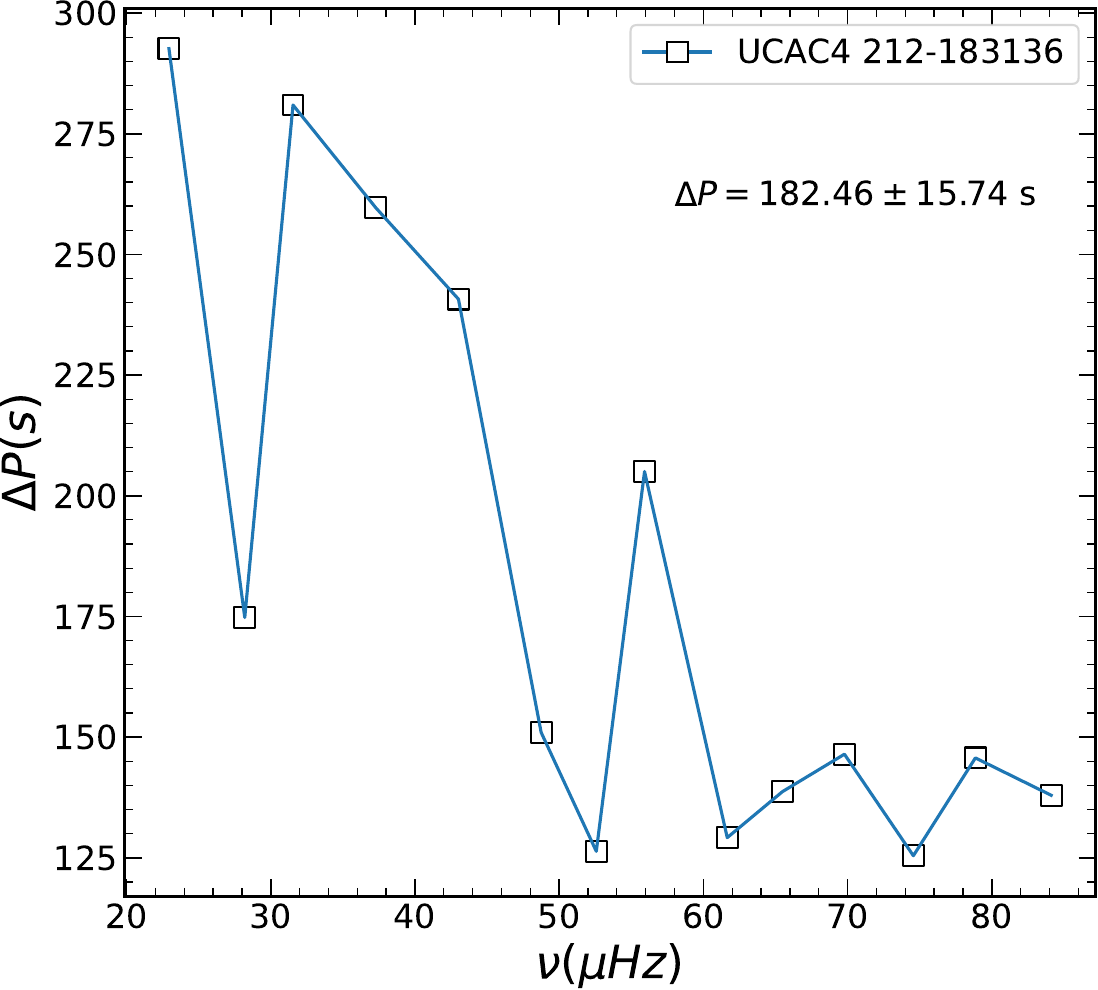} &
       \includegraphics[width=8cm]{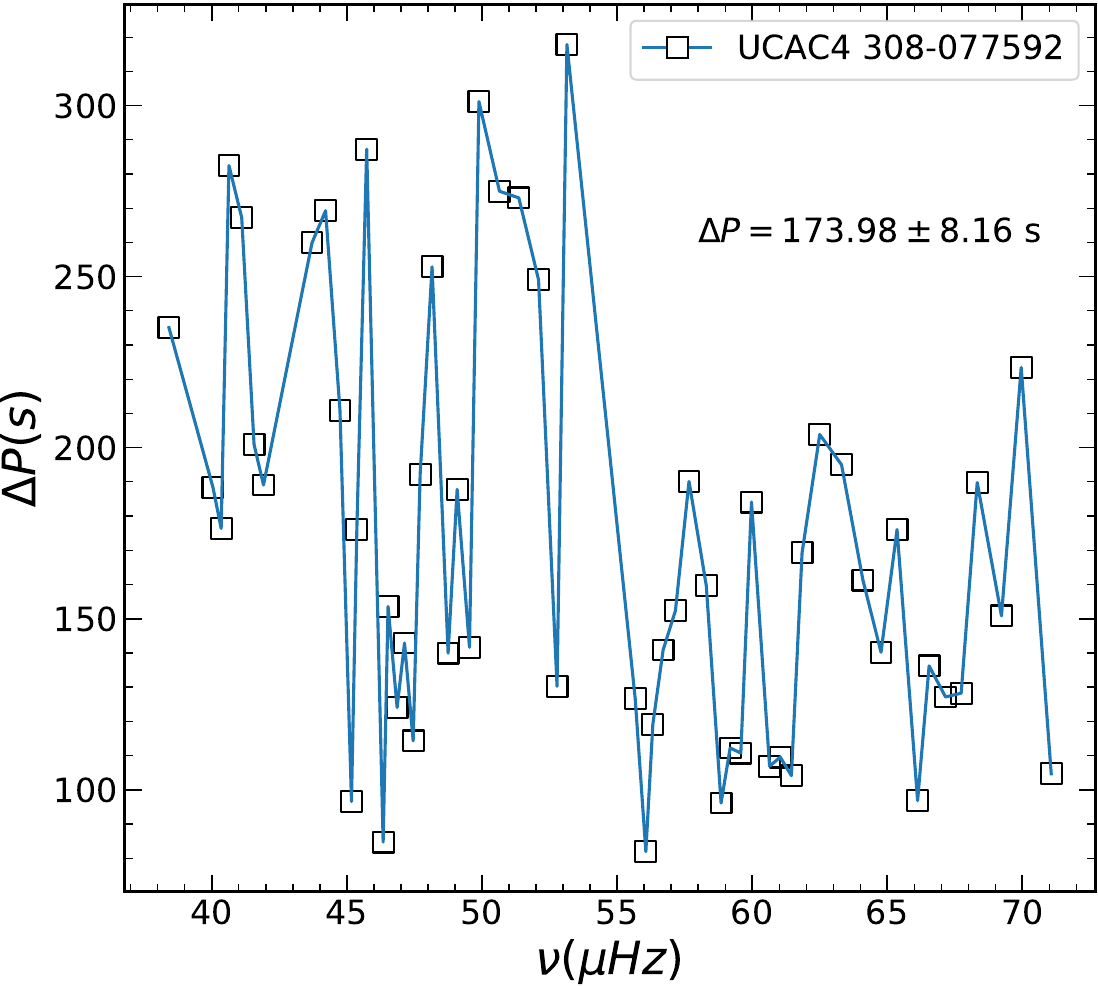} \\
       \includegraphics[width=8cm]{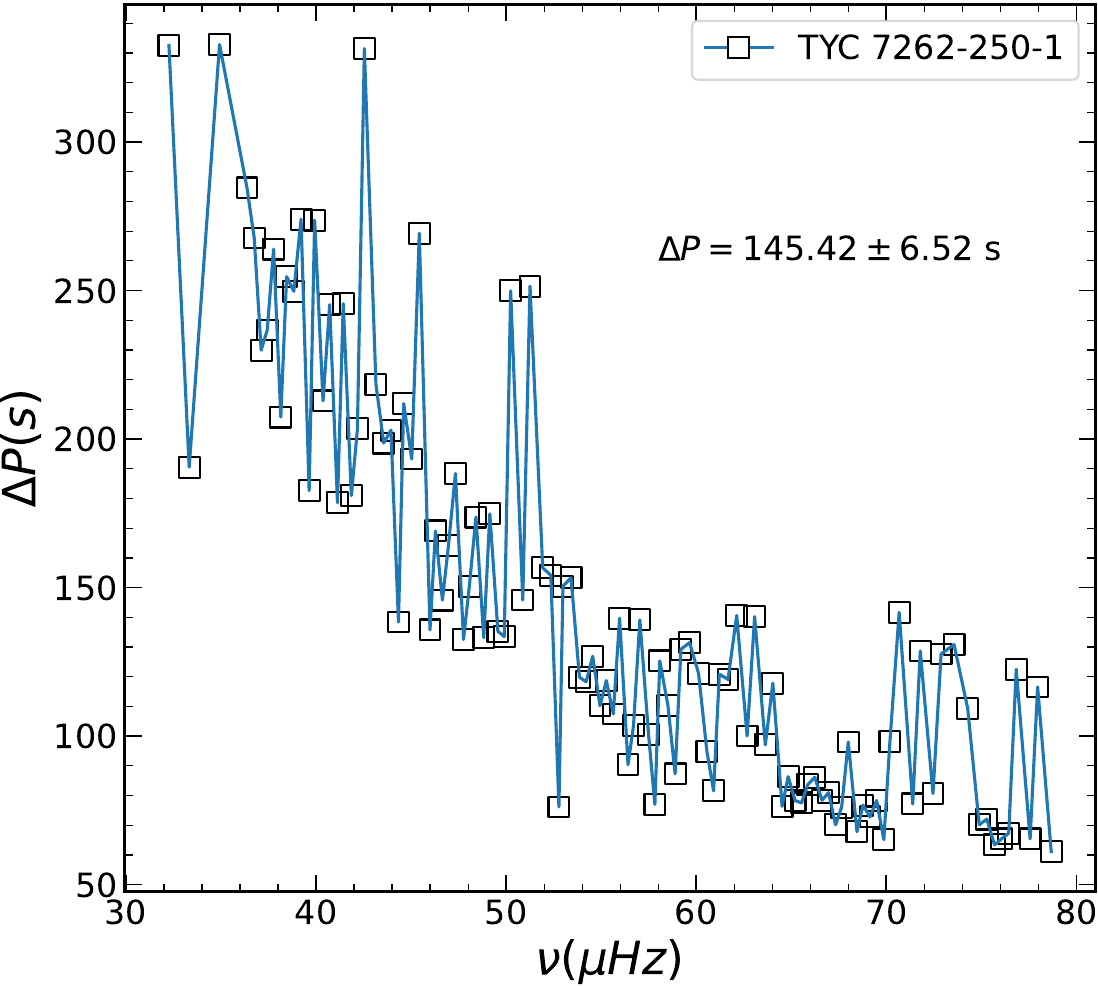} \\
    \end{tabular}
    \caption{Period spacing measurements for five SLR giants in this study.}
    \label{fig:period spacing}
\end{figure*}

We selected a small window in the background corrected PSD showing clear power excess, and the Auto-Correlation Function (ACF) is computed and integrated over this frequency range. The resulting collapsed ACF exhibits a distinct peak. 
The frequency value corresponding to the highest value of the collapsed ACF gives $\nu_{max}$. ACF convolves data with a lagged version of itself, and when oscillation modes overlap each other, spikes are seen in the ACF. Empirical relation given by \citet{Stello2009} gives a rough estimate of $\Delta\nu$

\begin{equation}
\Delta\nu = (0.263 \pm 0.009) \nu_{\text{max}}^{(0.772 \pm 0.005)} \, \mu\text{Hz}  
\end{equation}
Lightkurve identifies the peak of the ACF nearest to this estimate and reports it as $\Delta\nu$.  Owing to the shorter baseline time-series data provided by TESS, the signal-to-noise ratio (S/N) is comparatively low. Hence for the detection of significant oscillation peaks, we have used only frequencies corresponding to the standard criterion of S/N $>$ 3-4 \citep{Breger1993, Li2019}. The same background-corrected PSD is used to identify several consecutive dipole modes and estimate the median period spacing $\Delta P$. Measurement of $\Delta$P of each star is demonstrated in Fig. \ref{fig:period spacing}

\begin{figure}
\includegraphics[width=\columnwidth]{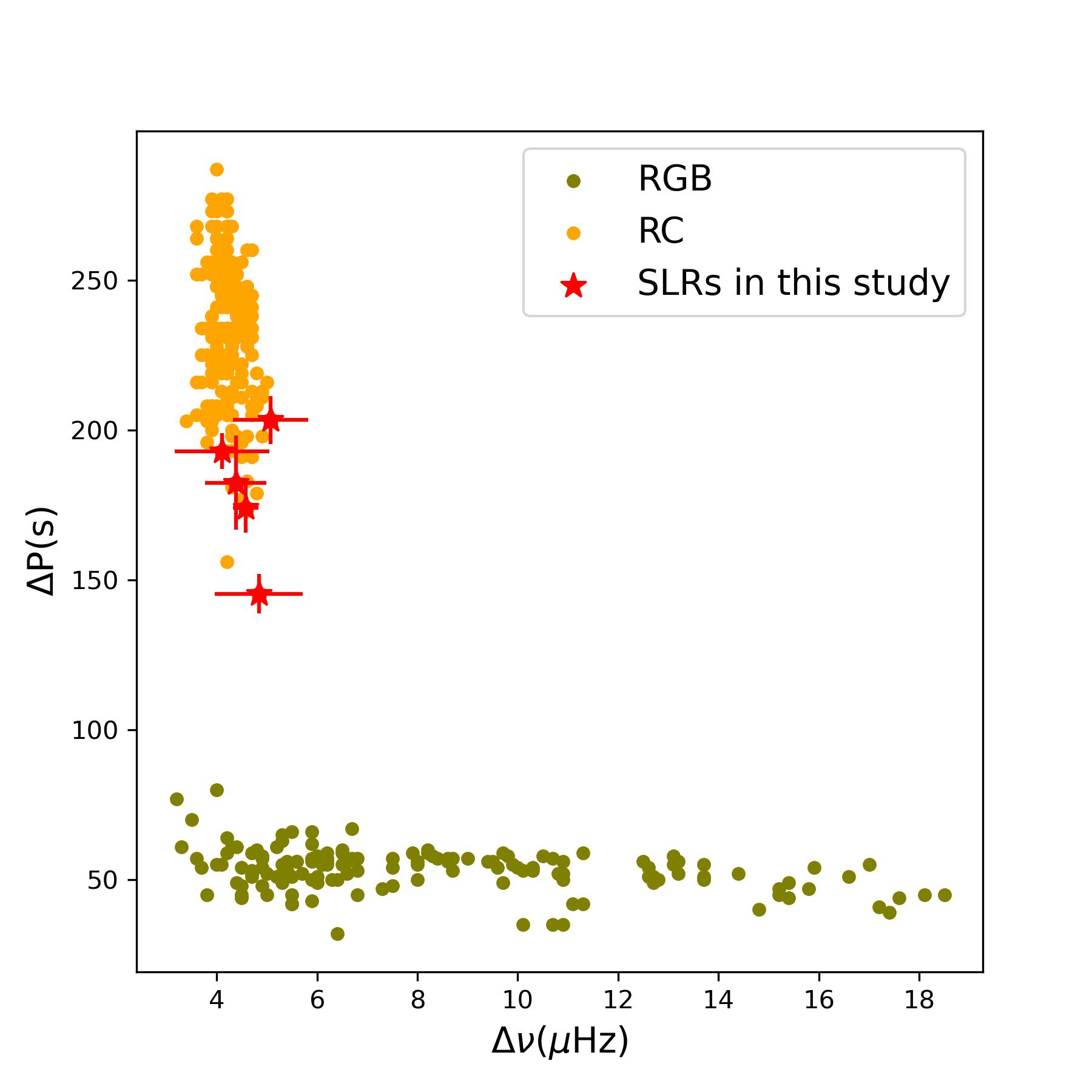}
\caption{ Asteroseismic plot of $\Delta$P - $\Delta\nu$. The dots with respective colours in the background represent stars classified according to evolutionary phases following the classification scheme given in \citep{bedding2011nature}. RGB stars are marked as olive circles, and RC stars are represented by orange circles. The five SLRs from this study (marked as red star symbol) show $\Delta$P and $\Delta\nu$ similar to RC stars.}
\label{fig:periods}
\end{figure}
Fig. \ref{fig:periods} shows the $\Delta$P-$\Delta\nu$ diagram of our sample. The RGB stars have a much denser core than core He-burning RC stars. Hence RC stars exhibit a more pronounced coupling in the mixed modes which leads to larger $\Delta$P values. Thus period spacing is a distinctive characteristic that provides a means to distinguish between these distinct phases of stellar evolution \citep{bedding2011nature}. We consider $\Delta$P $>$ 100s as the RC threshold \citep{Stello2013}.
The five giants are shown in  $\Delta$P-$\Delta \nu$ asteroseismic diagram. All five giants fall in the region occupied by giants of the post-He-flash phase with He-burning at the center, and none in the RGB region. Note, \citet{martell2021} list four of these in their catalog as Li-rich RGB giants using  stellar isochrones.
\par

{\bf Stellar parameters: mass, radius, log $g$ and Luminosity :} Another significant application of asteroseismology is the determination of fundamental stellar parameters such as mass, radius, and log $g$. They can be estimated using two seismic scaling relations\citep{Kjeldsen1995}: 
\begin{equation}
\frac{M}{M_\odot} \approx \left(\frac{\nu_{\text{max}}}{\nu_{\text{max},\odot}}\right)^3 \left(\frac{\Delta\nu}{\Delta\nu_\odot}\right)^{-4} \left(\frac{T_{\text{eff}}}{T_{\text{eff},\odot}}\right)^{3/2}   
\end{equation}
\begin{equation}
\frac{R}{R_\odot} \approx \left(\frac{\nu_{\text{max}}}{\nu_{\text{max},\odot}}\right) \left(\frac{\Delta\nu}{\Delta\nu_\odot}\right)^{-2} \left(\frac{T_{\text{eff}}}{T_{\text{eff},\odot}}\right)^{1/2}
\end{equation}
\begin{equation}
\frac{g}{g_\odot} \approx \left(\frac{\nu_{\text{max}}}{\nu_{\text{max},\odot}}\right)\left(\frac{T_{\text{eff}}}{T_{\text{eff},\odot}}\right)^{1/2}
\end{equation}
From Stefan-Boltzmann law, luminosity can be approximated as 
\begin{equation}
\frac{L}{L_\odot} = \left(\frac{R}{R_\odot}\right)^2 \left(\frac{T_{\text{eff}}}{T_\odot}\right)^4
\end{equation}.
\begin{deluxetable*}{cccccccc}
\caption{Asteroseismic parameters of sample giants\label{tab:seismic}}
\tablehead{
\colhead{Object name}&\colhead{$\nu_{max}$}&\colhead{$\Delta\nu$}&\colhead{$\Delta$P}&\colhead{Mass}&\colhead{Radius}&\colhead{log $g$}&\colhead{$\log\left(\frac{L}{L_{\odot}}\right)$}\\
\colhead{}&\colhead{($\mu$Hz)}&\colhead{($\mu$Hz)}&\colhead{(s)}&\colhead{(M$_{\odot}$)}&\colhead{(R$_{\odot}$)}&\colhead{(dex)}&\colhead{}}
\startdata
UCAC4 253-045343&36.57$\pm$ 1.09&4.11$\pm$0.94&194.99$\pm$5.90&1.17$\pm$0.53&10.84$\pm$3.03&2.44$\pm$0.01& 1.47$\pm$0.28\\
UCAC4 099-098976&40.62$\pm$2.17&5.08$\pm$0.75& 203.34$\pm$8.02& 1.04$\pm$0.24& 9.11$\pm$ 1.38&2.54$\pm$0.01&1.76$\pm$0.17\\
UCAC4 212-183136 & 31.55$\pm$1.12 & 4.38$\pm$0.61 & 182.46$\pm$15.74& 0.57$\pm$0.12& 8.14 $\pm$ 1.11& 2.37$\pm$0.01 &1.22$\pm$ 0.17\\
UCAC4 308-077592&45.62$\pm$3.39&4.58$\pm$0.25&173.98$\pm$8.16&1.40$\pm$0.04&10.62 $\pm$ 0.34 &2.53$\pm$0.01& 1.46$\pm$ 0.09\\
TYC 7262-250-1&37.99$\pm$1.69&4.84$\pm$0.88&145.42$\pm$6.52&0.72$\pm$0.18&8.31$\pm$ 1.33&2.46$\pm$0.01& 1.29$\pm$ 0.19\\
\enddata
\end{deluxetable*}
The solar reference values are $\nu_{\text{max},\odot}$ = 3090 $\mu$Hz, $\Delta\nu_\odot$ = 135.1 $\mu$Hz \citep{Huber2010} and $T_{\text{eff},\odot}$ = 5777.2K \citep{Prsa2016}
Although scaling relations provide efficient estimation of these parameters, greater accuracy can be achieved through asteroseismic grid modelling. This becomes more pertinent when examining metal-poor stars ([Fe/H] $<$ -1), as scaling relations tend to yield mass estimates that are inflated by roughly 16\% \citep{Epstein2014}. In this work, we have used grid modelling to measure mass, radius and log $g$. Grid modelling was performed using version 0.0.6 of the ASFGRID code \citep{Sharma2016,Stello2022ASFGRID}. Derived values for all stars are provided in Table \ref{tab:seismic}.
The log $g$ values from the asteroseismic analysis are significantly higher for giants compared to values derived from both the  Gaia parallax and in this study except for one star which is at RC, where all the methods yield the same log$g$ or the luminosity. 
\section{Discussion}\label{sec:discussion}

\begin{figure*}[!ht]
\centering
\includegraphics[width = \textwidth]{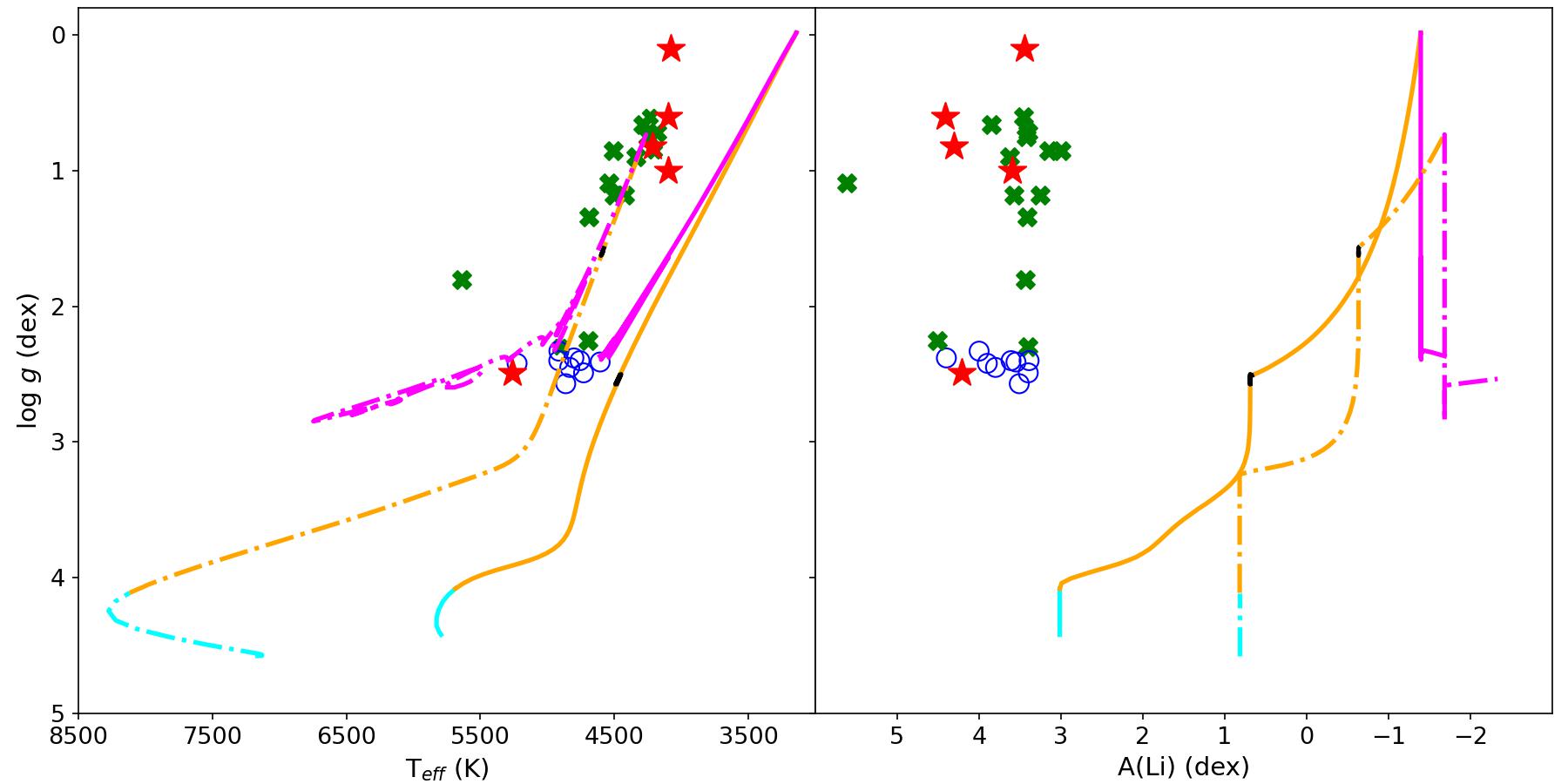}
\caption{ The five metal-poor SLRs in this study (star symbols) are shown in the left panel along with the known metal-poor SLRs ( crosses) from \citep{sitnova2023MNRAS,  kowkabany2022arxiv, sanna2020aa,hainingli2018, yan2018nature, kirby2012, ruchti2011lirich, kraft1999ApJ}, metal-rich SLRs ( circles) from \citep{Singh2021ApJ}, and two representative evolutionary tracks for [Fe/H]=-2.5 (dot-dashed line) and [Fe/H]=0.0 (continuous line). The colour code of the tracks indicates the evolutionary phases; core H-burning (cyan), shell H-burning ( orange), core He-burning, (magenta), and RGB luminosity bump (black). In the right panel, the same sample is shown along with the predicted MESA Li evolutionary models for two representative [Fe/H] values. The colour code on model tracks is the same as in the left panel.  } 
\label{fig:slrhrfeh}
\end{figure*}

The new metal-poor SLRs, along with the known SLRs, are shown in a plot of log $g$ and $T_{\rm eff}$ (Fig.  \ref{fig:slrhrfeh}: left panel). We have also shown a few metal-rich SLRs \citep{Singh2021ApJ}, which have been classified as red clump giants using asteroseismic analysis. Two metal-poor SLRs (one from our study and the other from \citet{hainingli2018}) fall in the RC region, and the rest are much brighter, overlapping with the early AGB space in the HR diagram.   We lack asteroseismically determined evolutionary phases for the already known metal-poor SLRs from the literature (see Fig \ref{fig:periods}). Since all five in this study are in the core-He-burning phase, similar to many SLRs in the metal-rich regime, it is reasonable to assume all the known metal-poor SLRs are also in the post-He-flash phase, and none are in the RGB phase. 
The same sample is shown in the right panel of Fig \ref{fig:slrhrfeh}, i.e., in a plot of log $g$ and A(Li), along with standard model predictions for Li abundances computed using MESA stellar evolutionary code (Modules for Experiments in Stellar Astrophysics) for 1M$_{\odot}$ with two representative values of [Fe/H]. As shown in the figure, Li only gets depleted as giants ascend the RGB, and the value reaches as low as A(Li) = -2.0~dex for metal-poor giants on reaching the red clump phase. This implies that the existence of SLRs at evolved phases, either at  RC or at the early AGB, does not comply with the existing theoretical models. 

The sample SLRs in Fig.\ref{fig:slrhrfeh} fall into two distinctive groups: one at RC near log $g$ $\sim$ 2.5 and the other group at higher luminosity, log $g$ $\sim$ 1.0. As per metal-rich Li-rich giants at RC, there are multiple evidences that Li enrichment occurs during the He-flash as the He-flash is the only major stellar event before stars arrive on RC  \citep{singh2019ApJ, kumar2020, Singh2021ApJ, Mallick2023}. It is also found that the Li-richness among giants is a transient phenomenon \citep{Singh2021ApJ} as Li starts depleting while they are on RC, which explains why the SLRs are rare. 
The 2nd group of SLRs at higher luminosity, beyond the RC phase, is interesting. The origin of high Li in these stars is not understood, but their existence suggests multiple mechanisms may be responsible for high Li in giants at different evolutionary phases. Here, we discuss  three possibilities for the 2nd group in the HR diagram: \\

1. One hypothesis could be that these giants are transitioning to RC post-He-ignition at the RGB tip. The location of stars in the HRD and their very high Li abundances do not rule out this possibility. The very high Li and the asteroseismic parameter, the average period spacing ($\Delta P$), imply that these giants have undergone He-flash very recently, which has been well demonstrated in the case of metal-rich Li-rich giants \citep[see][]{Singh2021ApJ}. If the He-flash is the universal origin of high Li in giants, these giants must be in the transition to RC. 
Other than these, we do not have any other observable signature to distinguish a giant, whether it is descending to the RC or evolving off from the RC phase as the evolutionary paths for a star descending towards RC and ascending towards the early AGB overlap in the HR diagram.  \\

However, given the time scales involved, the likelihood of finding stars during the transition may be very small. The giants are expected to take about 2 Million years to fully convert the degenerate He-core into a convective He-core burning phase of RC.   It is unclear whether the main He-flash causes the giants' sudden drop in luminosity to RC level. If so,  we expect the transition period to be about a few hundred days. The subsequent sub-flashes continue in stars at RC, converting the degenerate core into a fully convective core He-burning. Here, we can invoke the analogy of a typical Supernovae Ia (SN Ia) from low-mass stars whose light curves show a drop in brightness by a factor of 10-20 from their peak brightness within about 100 days \citep{pastorello2007mnras, pignata2008mnras}. The only difference is that the star gets disintegrated in the case of an SN explosion, and in the case of He-flash, the star is intact except for a sudden drop in the star's luminosity to RC level due to a shrink in the star's size. We ran the MESA model of 1M$_{\odot}$ with [Fe/H] = -2.5 and examined its evolution from the tip of RGB to the RC. We found that the transient period is about 500 days. Taking an SLR phase of about  2 Myrs (see \citep{Singh2021ApJ}), we find that the probability of finding SLRs during the transition to RC is about one in 2 Myrs. In the case of giants, \cite{deepak2019MNRAS} found 20 SLRs among the 51,982 RC giants, which is about one in 2500 or 0.04$\%$ probability. However, none were found during the short transition phase from the RGB tip to RC.  On the other hand, we found in this study four SLRs out of a total sample of 1038 metal-poor giants, i.e. one in 250 giants or 0.4$\%$, which is much larger than the expected probability of detecting SLRs during the transition period or even a factor 10 larger than SLRs among  RC giants. The much larger percentage of SLRs among metal-poor giants implies that these four giants may not be in the transition to RC. The higher percentage of  SLR among metal-poor giants also indicates SLR phase may last for longer among metal-poor giants compared to their counterparts in the metal-rich regime. Or Li-rich origin among metal-poor giants may differ from metal-rich giants. 

2. The second possibility is that the giants evolved from the RC phase to early AGB, with a carbon-oxygen core surrounded by He- and H-burning shells. Probably they have evolved from RC, and their high Li abundance may have a different origin other than the He-flash. Unfortunately, we do not have asteroseismic signatures for differentiating giants at RC from those that evolved off to the AGB phase. The normal C and $s$-process elements (see Table \ref{tab:abundstparam}), signatures of 3rd-dredge-up,  indicate the giants are still at the early AGB phase and yet to undergo 3rd-dredge-up. It is not clear how these giants became enriched with Li. Is the high Li inherited from the RC or produced $in situ$ at the early AGB? Indeed, the high Li cannot be from the RC phase, where most Li produced during the preceding He-flash event might have been destroyed. \citet{Singh2021ApJ} demonstrated that Li gets depleted rapidly and set a conservative upper limit of 40 Myrs for the SLR phase to last since the He-ignition began at the RGB tip. They argued that since only about 0.3 to 0.5 $\%$ of SLRs among RC giants, Li must be depleted rapidly. Thus,  the high Li in the 2nd group may have a different origin other than the He-flash. Many studies argued that the high Li among low-mass giants could be due to cool bottom processing (CBP) \citep{sackmann1999, ruchti2011lirich} in which additional mixing could occur between radiative layers beneath the deep convective envelope reaching by-products from H-burning shell.  Though the physical mechanism of such CBP is poorly understood, it was invoked for high Li seen in low-mass early AGB  stars \citep{abia2000ApJ}.
\\

3. The third possibility is the external origin, like the merger scenario explored by \cite{izzard2007,zhang2013}  to explain high Li among low mass R- or J-type stars.  \citet{zhang2013}  suggests that the merger with He white dwarf (HeWD) could produce single stars with high Li and similar luminosity to those in the 2nd group. The merger models do not expect enhancement of $s$-process elements and also no enhancement in carbon if the merger involved a red giant and low mass HeWD. \citet{piersanti2010aa} conducted a three-dimensional smoothed particle hydrodynamics (3D SPH) simulation of a merger between a low-mass HeWD and an RGB star, and they identified inefficient helium burning, resulting in no carbon enhancement. \citet{zhang2013} also obtained comparable outcomes from their one-dimensional post-merger calculations using low-mass HeWD models. The study by \citet{zhang2020ApJ} 
for a wide range of progenitors, mass binaries revealed that the post-merger abundances are a function of the mass of HeWDs. They postulated that Li-rich giants could form from mergers of low-mass HeWDs, (0.35M$_{\odot}$ $\le$ M$_{WD}$ $\le$0.40  M$_{\odot}$) with a low-mass RGB star. Evidence for stellar merger could be the presence of IR excess and dust. Unfortunately, only NIR data for these stars in the $2MASS$ JHK$_s$ bands is available, and their spectral energy distributions (SEDs) exhibit no IR excess. Far-infrared data is essential to confirm IR excess. \\

Interestingly, we found that luminosity values derived from asteroseismic analysis for the four SLRs are significantly lower than those found from Gaia parallaxes. However, for the RC giant UCAC4 099-098976, the luminosity values derived from both methods are the same (see  Table \ref{tab:luminMv} and Table \ref{tab:seismic}). We also checked the luminosity values derived from both methods for normal giants with little or no Li at the early-AGB and found no difference in the luminosity values. It is unclear at this point whether the difference could be due to the merger history of these SLRs. A merger could manifest the giant with higher luminosity due to increased mass and size. The increased mass may not affect the dense degenerate core at the centre; hence, there is little or no change in oscillations from the pre-merger giant. This implies that the four SLRs at higher luminosity probably had a merger history and not the one at RC. Unfortunately, the absence of TESS/Kepler data hindered our ability to confirm the evolutionary origin of two SLRs studied by {\citet{ruchti2011lirich} — whether they result from the evolution of single stars or involved a merger event.  

\section{Conclusion}\label{sec:conclusion}
In this study, we analysed photometric, spectroscopic and asteroseismic data of five Super Li-rich giants.  Among the five, one star is in the red clump phase, whereas the other four lie at higher luminosity near the early AGB. By comparing the existing metal-poor giants, we found two distinct groups of metal-poor SLRs: one at RC luminosity and the other at high luminosity near the early AGB. The SLRs at RC most likely originated from the He-flash, similar to metal-rich SLRs. The origin of SLRs at early-AGB may be either $insitu$ through cool bottom processing or external merger-induced nucleosynthesis and dredge-up. The significant difference between the luminosity values derived from Gaia parallaxes and the asteroseismology indicates some kind of merger events for the SLRs at higher luminosity values. We also discussed whether these giants are in the transition to the RC, post He-flash, and found it is unlikely as the observed SLR percentage among the metal-poor giants is much higher compared to the expected SLRs during the transition or among RC giants. 

Being the poorly explored population, large statistically uniform samples are required to better understand the evolutionary phase and Li production. The metal-poor Li-rich giants are crucial for understanding how the metallicity affects the Li production in giants. Also, it is important to measure other key chemical abundance ratios 
such as $^{12}C/^{13}C$ and C/N to understand the dredge-up process and evolutionary phase. 
\section{acknowledgments}
This work made use of the Third Data Release of the GALAH Survey \citep{buder2021}. The GALAH Survey is based on data acquired through the Australian Astronomical Observatory, under programs: A/2013B/13 (The GALAH pilot survey); A/2014A/25, A/2015A/19, A2017A/18 (The GALAH survey phase 1); A2018A/18 (Open clusters with HERMES); A2019A/1 (Hierarchical star formation in Ori OB1); A2019A/15 (The GALAH survey phase 2); A/2015B/19, A/2016A/22, A/2016B/10, A/2017B/16, A/2018B/15 (The HERMES-TESS program); and A/2015A/3, A/2015B/1, A/2015B/19, A/2016A/22, A/2016B/12, A/2017A/14 (The HERMES K2-follow-up program). We acknowledge the traditional owners of the land on which the AAT stands, the Gamilaraay people, and pay our respects to elders past and present. This paper includes data that has been provided by AAO Data Central (\url{https://datacentral.org.au}). All lightcurves for seismic analysis were obtained from the public data archive at MAST (\url{https://mast.stsci.edu/portal/Mashup/Clients/Mast/Portal.html}). This research made use of the open-source python packages Lightkurve \footnote{\url{https://github.com/lightkurve/lightkurve}} \citep{2018Lightkurve} and pySYD \footnote{\url{https://github.com/ashleychontos/pySYD}} \citep{2022Chontos}. This work has also made use of data from the European Space Agency (ESA) mission
{\it Gaia} (\url{https://www.cosmos.esa.int/gaia}), processed by the {\it Gaia}
Data Processing and Analysis Consortium (DPAC,
\url{https://www.cosmos.esa.int/web/gaia/dpac/consortium}). Funding for the DPAC
has been provided by national institutions, in particular, the institutions
participating in the {\it Gaia} Multilateral Agreement.
This research has made use of the SIMBAD database,
operated at CDS, Strasbourg, France. \citep{Wenger2000aa}. This research also has made use of the NLTE data obtained from the INSPECT database, version 1.0 (www.inspect-stars.net) \footnote{\url{http://www.inspect-stars.com/}}
\appendix
\section*{Lightkurve results for sample giants}
\begin{figure}[H]
\centering
\begin{tabular}{cc}
\includegraphics[width=0.54\textwidth]{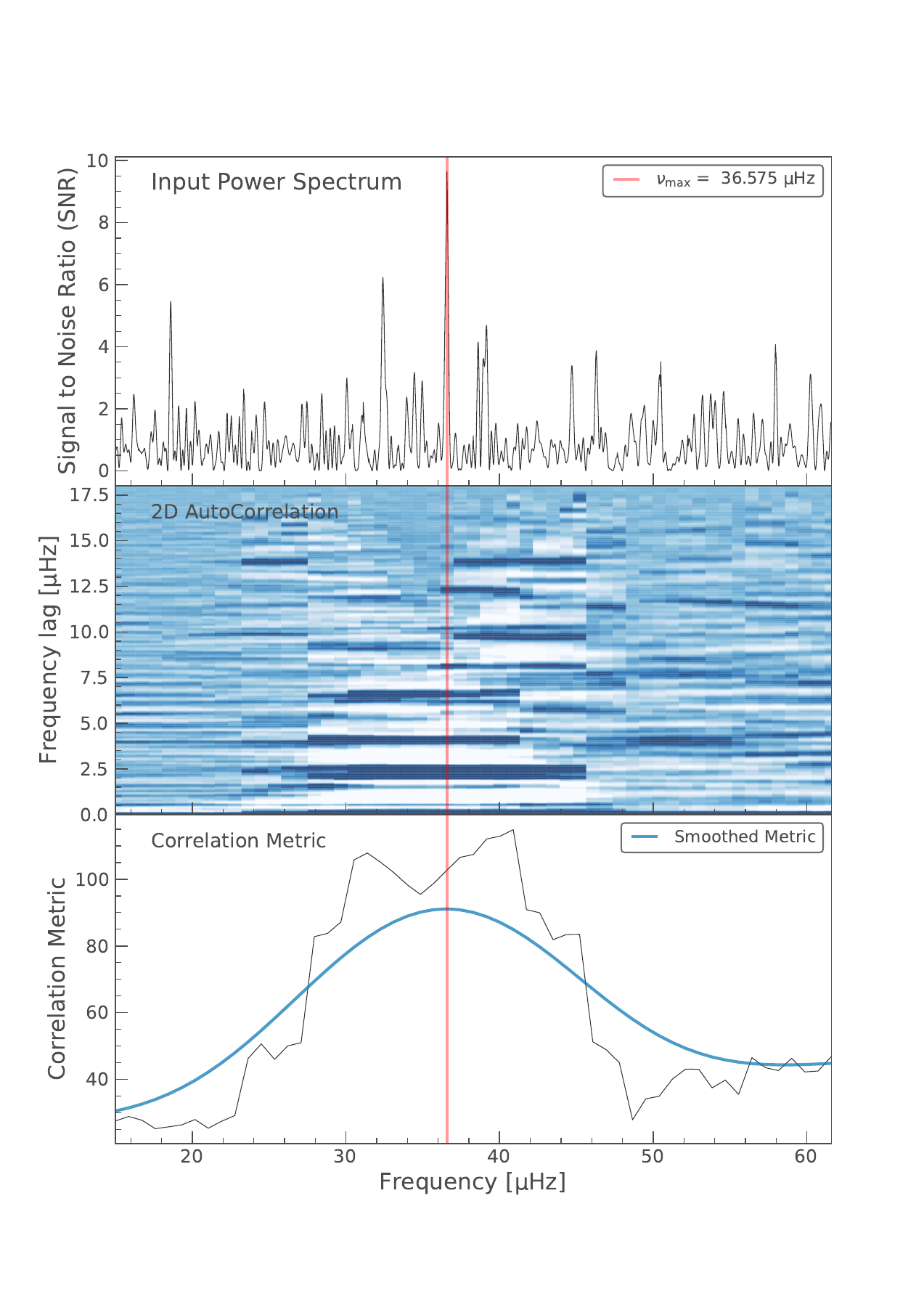} &\includegraphics[width=9.2cm,height=13.8cm]{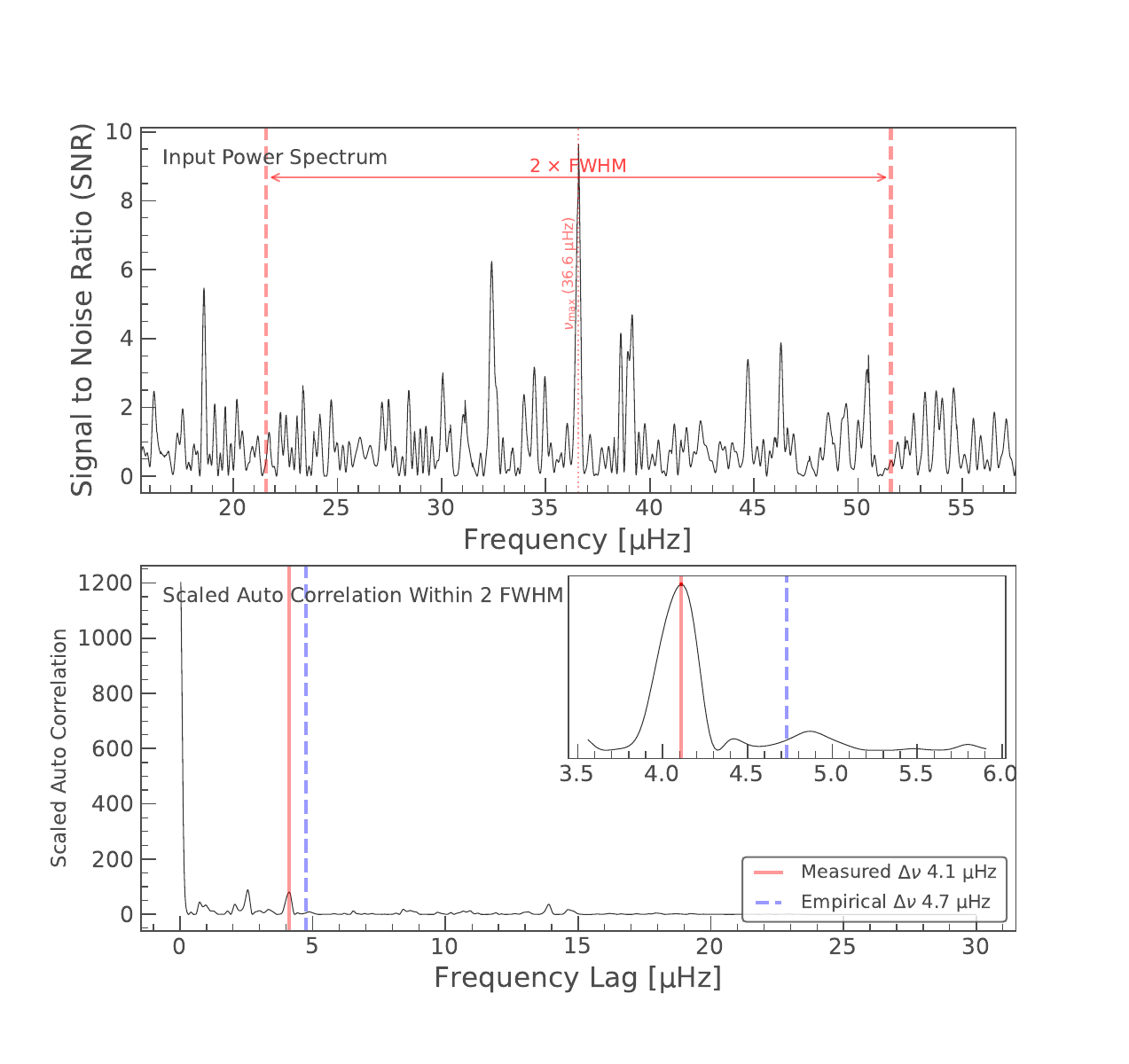}  \\
\end{tabular}       
\caption{UCAC4 253-045343}
\label{fig:star1}
\end{figure}
Results of Lightkurve analysis of the star UCAC4 253-045343. The left subplot has 3 panels -  The top panel displays the signal-to-noise ratio (SNR) periodogram. The vertical red line indicates $\nu_{max}$. Middle Panel: In the middle panel, we apply a two-dimensional ACF to different segments of the periodogram. Lower Panel: In the bottom panel, we present the mean collapsed ACF as a function of the central frequency of each segment. A Gaussian curve (shown in blue) is the smoothened collapsed ACF. The right subplot has two panels - The upper panel displays the frequency region over which ACF is evaluated. In the lower panel, we present the ACF itself. This is the outcome of computing the correlation between the data and itself while it is progressively shifted over itself.
\bibliography{slr}{}
\bibliographystyle{aasjournal}

\end{document}